\documentclass[12pt,preprint]{emulateapj}

\def\solmass {\,$\hbox{M}_\odot$}
\def\solum {\,$\hbox{L}_\odot$}
\def\kms {\hbox{\,${\rm km\, s}^{-1}$}}
\def\water {H$_2$O}
\def\meth {CH$_3$OH}
\def\chcn {CH$_3$CN}
\newcommand{\hii}{\ion{H}{2}}
\newcommand{\sii}{\ion{S}{2}}

\newcommand{\ts}{\thinspace}
\newcommand{\simless}{\mathbin{\lower 3pt\hbox
     {$\rlap{\raise 5pt\hbox{$\char'074$}}\mathchar"7218$}}}
\newcommand{\simgreat}{\mathbin{\lower 3pt\hbox
     {$\rlap{\raise 5pt\hbox{$\char'076$}}\mathchar"7218$}}}

\newcommand{\aboutless}{$\simless$\ts}

\shorttitle{CARMA Masers}
\shortauthors{Schnee \& Carpenter}

\begin{document}

\title{Testing the Evolutionary Sequence of High Mass Protostars with
CARMA} 

\author{Scott Schnee\altaffilmark{1} \& John M. Carpenter\altaffilmark{1}}

\email{schnee@astro.caltech.edu}

\altaffiltext{1}{Department of Astronomy, California Institute of
Technology, MC 105-24 Pasadena, CA 91125}

\begin{abstract}

We present 1\arcsec\ resolution CARMA observations of the 3\,mm
continuum and 95\,GHz \meth\ masers toward 14 candidate high mass
protostellar objects (HMPOs).  Dust continuum emission is detected
toward seven HMPOs, and \meth\ masers toward 5 sources. The 3\,mm
continuum sources have diameters $<$ $2\times 10^4$AU, masses between
21 and 1200\solmass, and volume densities $> 10^8$\,cm$^{-3}$.
Most of the 3\,mm continuum sources are spatially coincident with
compact \hii\ regions and/or water masers, and are presumed to be
formation sites of massive stars. A strong correlation
exists between the presence of 3\,mm continuum emission, 22\,GHz
\water\ masers, and 95\,GHz \meth\ masers. However, no 3\,mm continuum
emission is detected toward ultracompact \hii\ regions lacking maser
emission. These results are consistent with the hypothesis that 22\,GHz 
\water\ masers and \meth\ masers are signposts of an early phase in the 
evolution of an HMPO before an expanding \hii\ region destroys the 
accretion disk.

\end{abstract}

\keywords{stars: formation  -- masers}

\section{Introduction}

The evolutionary sequence of low mass star formation is broadly
understood.  Gravity in a prestellar core eventually overcomes the
thermal, turbulent, magnetic and rotational support to form a protostar.
This protostar is surrounded by an envelope and accretion
disk, and launches a bipolar outflow.  As the protostar accretes mass,
the surrounding envelope and disk are cleared by the winds and
outflows until no more mass is left to accrete \citep[see,
e.g.,][]{McKee07}.

High mass protostars share many of the same observational
characteristics as their low mass counterparts: outflows
\citep{Zhang01}, jets \citep{Caratti08}, envelopes \citep{Molinari96}
and possibly disks \citep{Cesaroni05}.  However, the elevated
radiation environment that accompanies high mass stars raises the
question as to what degree high mass star formation is merely a scaled
version of low mass star formation.  Answering that question is not
straightforward, partly because high mass star forming regions are
less common than the low mass counterparts, and tend to be more distant.  
The greater distance, coupled with the propensity of HMPOs to form in
clusters and the disruptive influence of jets, outflows and \hii\
regions, complicates interpretation of the observations
\citep{Zinnecker07}.  Therefore, to study the initial conditions of
high mass star formation, it is important to identify young HMPOs
before their surroundings have been altered by energetic processes.

Surveys for high mass protostars have been conducted over the years at
a variety of wavelengths.  Optical and radio continuum surveys of the
galactic plane have identified \hii\ regions in which massive star
formation is already well underway \citep[e.g.][]{Marsalkova74,
Altenhoff79, Haynes79}. \citet{Wood89} used the {\it IRAS} catalog to
search for far-infrared sources associated with ultracompact \hii\
(UCHII) regions and high mass protostars embedded in dense cores.
Massive star formation is thought to be associated with Infrared Dark
Clouds (IRDCs), which are seen in absorption against the diffuse
mid-infrared background of the galactic plane \citep{Perault96,
Egan98}. The masses of IRDCs suggest that they may form stellar
clusters \citep{Rathborne06, Rathborne07}, and IRDCs are known to host
both protostars and starless cores \citep{Wang06}. More recently,
surveys with the {\it Spitzer} Space Telescope such as GLIMPSE
\citep{Benjamin03} have identified HMPOs based on mid-infrared colors
\citep{Kumar07,Cyganowski08}.

High mass star formation also coincides with a variety of masers, as
demonstrated by targeted surveys of known HMPOs and blind surveys of
the galactic plane \citep[e.g.][]{Szymczak00,Pandian07}.
Interferometric observations have shown that water masers (at
22.2\,GHz) are associated with J-shocks from jets and winds ejected by
massive YSOs \citep{Torrelles03, Goddi05, Moscadelli05}.  Methanol
masers (at 6.7\,GHz) are less understood, but are thought to trace the
shock where a jet meets the ambient circumstellar gas
\citep{DeBuizer03} and/or the innermost portions of a disk around a
protostar \citep{Edris05}.  Hydroxyl maser emission (at 1.7\,GHz) may
also come from a disk around an accreting protostar \citep{Edris05},
and is also associated with both \hii\ and UCHII regions
\citep{Edris07}.

Although there are many tracers of high mass star formation (masers,
radio continuum, infrared objects), they are not always spatially
coincident or even found within the same object.  Over the years,
various observations of HMPOs have been unified under a common
evolutionary sequence \citep[see reviews by][]{Garay99, Menten05,
Beuther07, Zinnecker07}. IRDCs are thought to host the youngest
protostars, perhaps surrounded by accretion disks and driving jets and
outflows.  The disk, jet, and outflow system are conducive for the
formation of \water, \meth, and possibly OH maser emission.  This
evolutionary stage is also distinguished by dust emission from the
disk and envelope around the protostar. The development of an \hii\
region is suppressed initially by the continued accretion of material
onto the star \citep{Walmsley95}. As the protostar increases in mass,
temperature, and luminosity, an \hii\ region forms that is initially
gravitationally bound to the star and remains compact in size
\citep{Keto02}. Eventually the protostar becomes massive enough to
produce an \hii\ region that escapes the stellar gravitational
potential \citep{Keto07}. The \hii\ region expands to engulf the
accretion disk and envelope, which destroys the source of the masers,
jets, and outflows \citep{Lo75, Genzel77, Codella95, Codella04}. This
evolutionary sequence has provided a framework to explain a multitude
of observations \citep[e.g.][]{Molinari96, Beuther02b, Minier05,
vanderTak05}.

An observational prediction of this evolutionary sequence is that
compact dust continuum emission from an accretion disk or envelope
will be present in HMPOs containing maser emission
\citep{Codella04}. At a more advanced stage when the \hii\ region has
grown, maser emission will be suppressed and dust emission from the
disk and envelope should be at a much reduced level. Single dish maps
of the dust emission exist for a number of HMPOs \citep{Beuther02a},
but no clear correlation is present between the dust continuum
luminosity and the presence of maser emission. However, the resolution
of these observations (11\arcsec) is insufficient to distinguish
compact emission around HMPOs from the ambient dense
core. Interferometric observations have achieved significantly higher
resolution ($\sim$5\arcsec), but they cover a limited range of presumed
evolutionary states and are insufficient to test the proposed
evolutionary sequence \citep{Molinari02, Beltran04, Cesaroni05,
Zhang07, Reid08}.

In this paper, we present high resolution (1\arcsec) CARMA
observations of the dust continuum and 95\,GHz \meth\ maser emission
toward 14 HMPOs to test the evolutionary sequence for massive stars
described above. The HMPO sample was selected to cover
the primary evolutionary stages in this model. With the order of
magnitude improvement in angular resolution over previous single dish
surveys, we can isolate any compact dust emission originating from a
disk or envelope associated with an HMPO. In \S\ref{SAMPLE}, we
describe the sample of HMPOs in our survey, and the available
ancillary information regarding their evolutionary state. The new
CARMA observations are described in \S\ref{OBSERVATIONS} and the
observational results are presented in \S\ref{RESULTS}. In
\S\ref{DISCUSSION}, we compare our results with the proposed
evolutionary sequence for massive stars.

\section{Sample} \label{SAMPLE}

We drew a sample of 14 high-mass protostellar candidates from the
survey of \citet{Sridharan02}.  The HMPOs in the \citet{Sridharan02}
list have {\it IRAS} colors of UCHII regions \citep{Wood89}, are
detected in the CS J=2--1 survey of \citet{Bronfman96}, are bright at
FIR wavelengths ($F_{60} > 90$ Jy, $F_{100} > 500$ Jy), are north of
$-$20\degr\ declination, and are not detected above 25 mJy in the 5\,GHz
radio continuum surveys of \citet{Gregory91}, \citet{Wright94} and
\citet{Griffith94}. 

The properties of the HMPOs selected for this study are summarized in
Table~\ref{ARCHIVALTAB}. The HMPOs have a bolometric
luminosity between $10^{3.9}$ and $10^{5.0}$\solum. The distances to the
sources range from 1.7 to 10.5\,kpc, where the near-far distance
ambiguity has been resolved through a variety of methods as described
in \citet{Sridharan02}.  The dense cores associated with the HMPOs
have masses of $\sim$400-10000\solmass\ contained within a $\sim$1~pc
region.  We adopted the velocity and association with cm-continuum
emission, \water\ masers and \meth\ masers for each source given in
\citet{Sridharan02}.  The 3.6\,cm radio continuum emission was
observed with the VLA in B-array with a sensitivity of $\sigma\sim$1
mJy, and the coordinates of the detections were kindly provided by
H.~Beuther (private communication).  The 1.2\,mm continuum fluxes and
positions were measured with MAMBO on the IRAM 30m telescope at a
resolution of 11\arcsec\ \citep{Beuther02a}.  The association with OH
masers is presented in \citet{Edris07}, and were detected with the
Nancay radio telescope and the GBT.  

The 14 HMPOs that we observed were divided into three categories
depending on the presence or absence of cm-continuum emission
indicative of an UCHII region, 6.7\,GHz \meth\ masers, or 22\,GHz
\water\ masers.  In order of increasing age according to the proposed
evolutionary sequence \citep{Lo75, Genzel77, Codella95}, our sample
contains (1) five sources with (\meth\ or \water) maser emission but
no cm-continuum, (2) five sources with both cm-continuum and maser
emission, and (3) four sources with cm-continuum but no maser
emission.  For each of these three categories, the range of distances
($2.9-9.5$\,kpc, $1.7-10.5$\,kpc, $2.6-5.5$\,kpc, respectively) and
luminosities ($10^{4.1}-10^{5.0}$\solum, $10^{3.9}-10^{4.9}$\solum,
$10^{4.1}-10^{4.7}$\solum) sampled are similar.  We centered each
CARMA map on the position of the peak 1.2\,mm continuum emission
derived from 11\arcsec\ resolution single-dish observations
\citep{Beuther02a}.  Note that in \citet{Sridharan02} the source IRAS
18345$-$0641 is associated with both maser and cm-continuum
emission. For the purposes of this survey, we consider IRAS
18345$-$0641 to be a maser-only HMPO because the masers and 1.2\,mm
continuum peak are coincident to within 1\arcsec, but the cm-continuum
emission is offset from the maser position by 44\arcsec, or 2~pc.

\section{CARMA Observations} \label{OBSERVATIONS}

Continuum and spectral line observations in the 3\,mm window were
obtained with CARMA (Combined Array for Research in Millimeter-wave
Astronomy).  CARMA is a 15 element interferometer consisting of nine
6.1 meter antennas and six 10.4 meter antennas.  Data were taken in
the CARMA B-array configuration between 2007 December 24 and 2008
February 11.  The projected baselines in this configuration range from
65m to 800m. The CARMA correlator records signals in three separate
bands, each with an upper and lower sideband.  Two bands were
configured for maximum bandwidth (468 MHz with 15 channels per band)
to observe continuum emission, providing a total continuum bandwidth
of 1.87\,GHz.  One band was configured with 31 MHz bandwidth across 63
channels (with a resolution of 0.488 MHz or 1.6 \kms\ per channel) to
observe the hyperfine lines of \chcn\ (5--4) in the lower sideband
(from 92.959 to 92.987\,GHz) and \meth\ (8$_{0,8}$--$7_{1,7}$) in the
upper sideband (at 95.2\,GHz).  The FWHM of the synthesized beam with
natural weighting of the visibilities is $\sim$1\arcsec\ (see
Table~\ref{OBSPARAM}), and the largest angular scale that can be
accurately imaged in the maps is $\sim$5\arcsec.  The half-power beam
width of the 10.4m antennas is 73\arcsec\ at the observed frequencies.

The observing sequence was to integrate on a primary phase calibrator
for 2.5 minutes, an HMPO for 7 minutes, and a secondary phase
calibrator for 2.5 minutes.  This cycle was repeated for $\sim$3 hours
for each HMPO.  Calibration and imaging were done using the MIRIAD
data reduction package \citep{Sault95}.  The sources observed, along
with the phase center, calibrators and the noise in the resultant
continuum maps and spectra are listed in Table~\ref{OBSPARAM}.  The
secondary calibrator was used to test the accuracy of the phase
transfer from the primary calibrator.  A passband calibrator (3C454.3)
was observed for fifteen minutes in each set of observations, and
radio pointing was performed every two hours thereafter.  Absolute
flux calibration was accomplished using Neptune, Uranus and MWC349 as
primary flux calibrators and 3C454.3 as a secondary flux calibrator.
Based on the repeatability of the quasar fluxes, the estimated random
uncertainty in the measured source fluxes is $\sigma\sim5$\%.  Because
we spent only $\sim$3 hours on each of the fourteen HMPOs we had
sufficient signal to noise to detect \chcn\ in only one source, IRAS
20126+4104, and the \chcn\ observations are not discussed further.

\section{Results} \label{RESULTS}

Here we present the CARMA 3~mm continuum and 95~GHz \meth\ maps and
compare them with maps of 6.7~GHz \meth\ and 22~GHz \water\ masers and
radio continuum emission taken from the literature.  We derive masses
for the 3~mm detections, and present evidence for outflows eminating
from these sources.

\subsection{3~mm continuum}

We detect 3\,mm continuum emission toward seven of the fourteen HMPOs
at the 3$\sigma$ level or greater.  Contour plots of the 3\,mm flux
density for the detections are presented in Figure~\ref{CONTOURPLOTS},
and plots of the non-detections are shown in
Figure~\ref{NONCONTOURPLOTS}.  The sources IRAS 23033+5951 and IRAS
19217+1651 are resolved into three and two components,
respectively. None of the detected sources are more than a factor of
$\sim$2 larger than the $\sim$1\arcsec\ beam of our maps. The
secondary phase calibrators, assumed to be point sources, have the
same range of sizes as the detected HMPOs, so all of the 3\,mm
detections are consistent with point sources at 1\arcsec\ resolution
blurred by seeing effects.  The CARMA observations are sensitive to
angular scales as large as 5\arcsec, and resolve out the majority of
the extended emission associated with the star-forming regions seen in
the 1.2~mm single dish maps \citep{Beuther02a}.

The positions and measured flux densities of the 3\,mm continuum
sources are listed in Table~\ref{OBSPROP}. The brightest 3\,mm
continuum source detected in each map is located within 6\arcsec\ of
the centroid of the larger scale core traced by single dish 1.2\,mm
continuum emission as measured by \citet{Beuther02a}. Given the
11\arcsec\ resolution of the 1.2\,mm observations and the difficulty
in finding the exact center of non-symmetric intensity distributions,
the offsets between the 3\,mm and 1.2\,mm positions are most likely
not significant.

\subsection{Masers}

We detect 95\,GHz \meth\ emission toward five HMPOs. Two of the
sources (IRAS 23033+5951 and IRAS 23151+5912) have a single \meth\
emission detection, and three (IRAS 18517+0437, IRAS 20126+4104 and
IRAS 23139+5939) have multiple detections.  Spectra for each of the
detections are shown in Figure~\ref{SPECTRAPLOTS}, and spectral
parameters are summarized in Table~\ref{METHPROP}.  The 95\,GHz \meth\
lines are bright (with brightness temperatures around a few hundred
Kelvin), narrow (unresolved at 1.6 \kms\ resolution) and spatially
unresolved. Therefore the inferred brightness temperatures are lower
limits and the lines are likely to be masing.

\begin{figure}
\epsscale{0.9}
\plotone{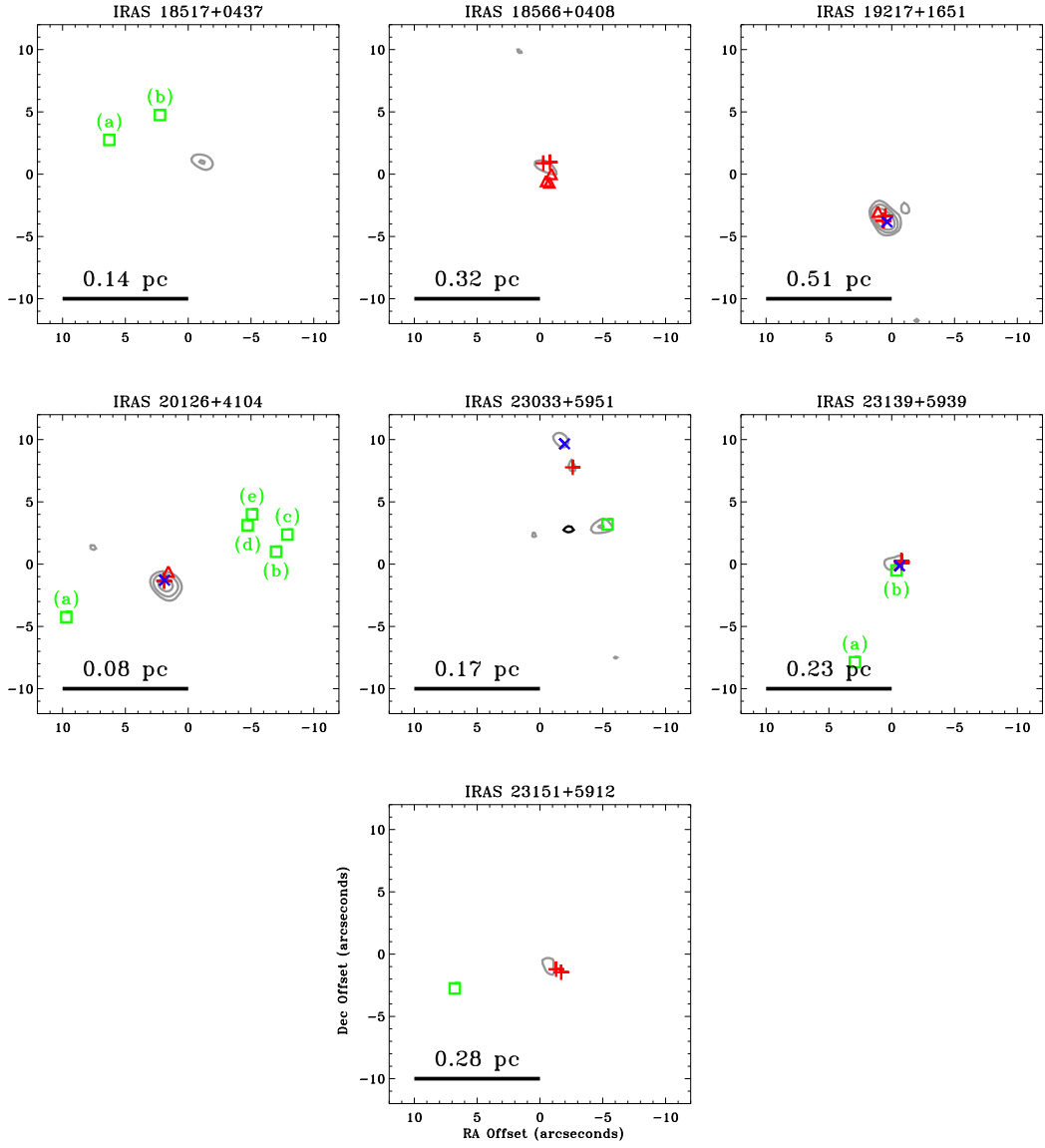}
\caption{Contour plots of the 3\,mm continuum flux density for HMPO
candidates sources detected with CARMA at the $\ge 3\sigma$ level,
centered at the position of the peak 1.2~mm MAMBO (11\arcsec\
resolution) emission.  Grey contours show emission at 3, 6 and 12
times the noise in each map, and black contours show emission at $-$3,
$-$6 and $-$12 times the noise.  Red crosses and triangles show the
positions of 22\,GHz \water\ and 6.7\,GHz \meth\ masers, respectively,
and blue 'X' mark the positions of 3.6\,cm continuum sources
\citep[][private communication with Beuther]{Beuther02b,Sridharan02}.
The source IRAS 18517+0437 was not observed by
\citet{Beuther02b}. Green squares show the 95\,GHz \meth\ maser
emission detected by CARMA. The resolution of the 3\,mm continuum
images is $\sim$1\arcsec.
\label{CONTOURPLOTS}}
\end{figure}

\begin{figure}
\epsscale{0.9} \plotone{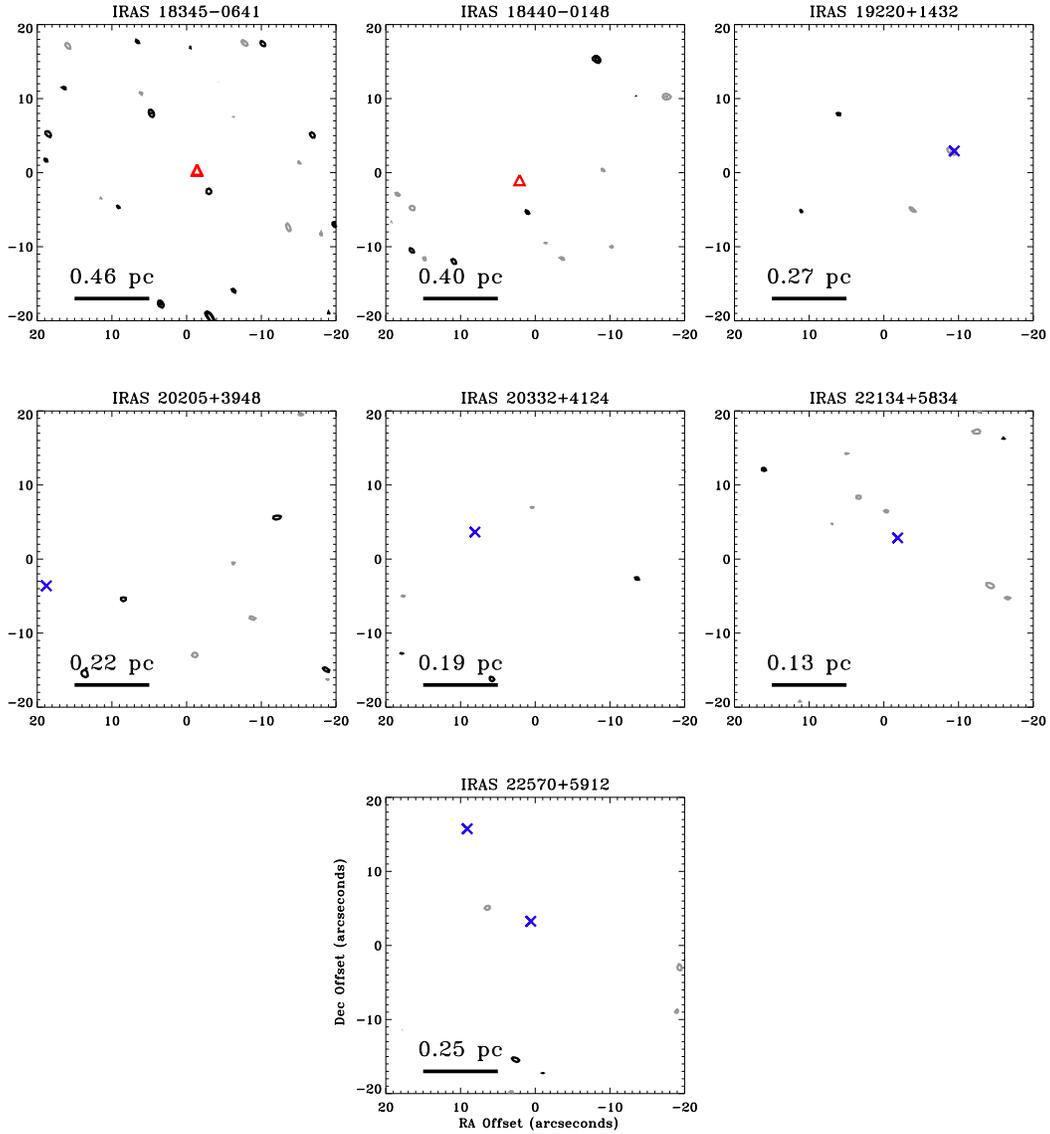}
\caption{Contour plots of the 3\,mm continuum flux density for eight
HMPO candidates not detected with CARMA.  Grey contours show fluxes
at 2.0 and 2.5 times the noise in each map, and black contours show
fluxes $-$2.5 and $-$2.0 times the noise.  The symbols are otherwise
the same as presented in Fig. \ref{CONTOURPLOTS}.
\label{NONCONTOURPLOTS}}
\end{figure}

\begin{figure}
\epsscale{1.0}
\plotone{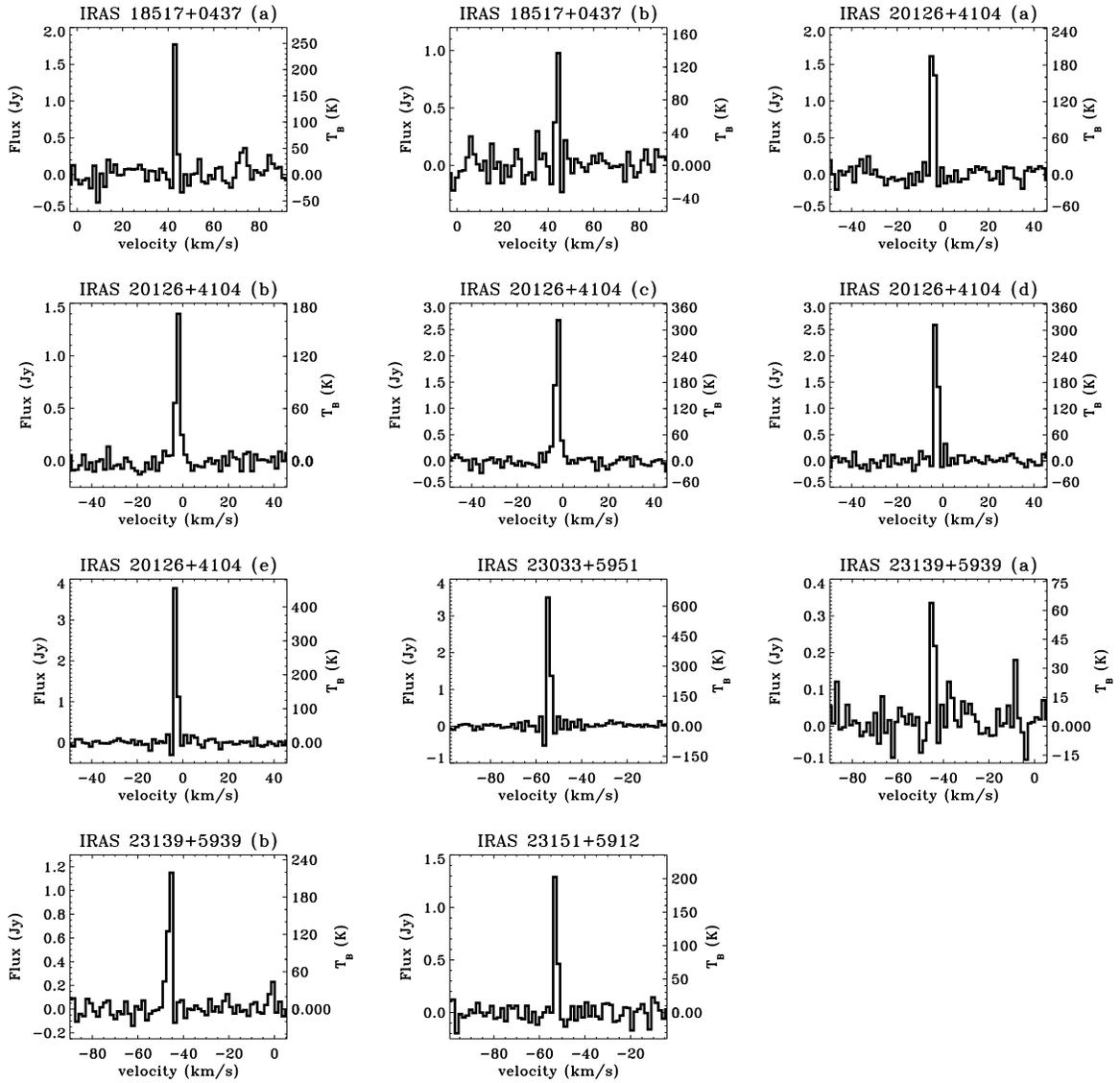}
\caption{Spectra of the 95\,GHz \meth\ maser detections, with the
same identifications shown in Figure~\ref{CONTOURPLOTS} and Table 
\ref{METHPROP}.
\label{SPECTRAPLOTS}}
\end{figure}

The five HMPOs that have 95\,GHz \meth\ masers are also detected in
the 3\,mm continuum.  The location of the presumed masers are marked
on the 3\,mm contour maps in Figure~\ref{CONTOURPLOTS}. Two 95\,GHz
\meth\ masers are coincident with the 3\,mm continuum emission to
within 0.6\arcsec\ (or $<$ 2000-3000\,AU), while the other nine masers
are separated by $\sim$10$^4$ AU from the 3\,mm continuum peaks (see
Table~\ref{METHPROP}).  The projected spatial offsets are consistent
with the expectation that the 95\,GHz masers are collisionally pumped
at the interface between outflows and the interstellar medium
\citep{Minier02}.  All of the \meth\ lines are found within 2\,\kms\
of the systemic velocity of the HMPO, implying that the masers form in
the colliding gas. Our finding that the maser line is found at nearly
the same velocity as the ambient medium is consistent with the results
of \citet{Minier02}, who find that this maser line is often, but not
always, at the systemic velocity of the HMPO.

The location of known centimeter continuum sources, \water\ masers,
and \meth\ masers are marked on the contour maps in
Figures~\ref{CONTOURPLOTS} and \ref{NONCONTOURPLOTS}. Six of the seven
CARMA 3\,mm continuum detections (all except IRAS 18517+0437) were
included in an interferometric survey of 22\,GHz \water\ and 6.7\,GHz
\meth\ masers by \citet{Beuther02b}.  Water masers have been
identified toward all six cores detected with CARMA, and three of
these cores also have 6.7\,GHz \meth\ masers.  For HMPOs with both
22~GHz \water\ and 6.7~GHz \meth\ masers, the two types of masers are
spatially coincident to within 1-2\arcsec. Both types of masers also
fall within $\sim$1\arcsec\ of our 3\,mm continuum positions, as shown
in Figure~\ref{CONTOURPLOTS}.  Of the multiple 3\,mm components
resolved in IRAS 19217+1651 and IRAS 23033+5951, one is coincident
with \water\ masers in each of the HMPOs.

\subsection{Radio Continuum}

Four of ten 3\,mm continuum sources are located within 1\arcsec\ of a
3.6\,cm radio continuum source. The 3\,mm continuum emission for these
objects may originate from free-free or thermal dust emission.  If we
assume that the sources in our sample are typical of other UCHII
regions, the free-free emission will be optically thick (i.e. $S_\nu
\propto \nu^2$) for $\nu < 10-15$\,GHz, and optically thin ($S_\nu
\propto \nu^{-0.1}$) for higher frequencies \citep{Hunter98, Turner04,
  Zapata08, Hunter08}.  For a ``typical'' UCHII radio continuum
spectrum, the expected 3\,mm continuum flux density from free-free
emission will be less than a factor of 2.7 higher than the 3.6\,cm
flux density.  We estimate the relative contributions of free-free
and dust emission to the observed 3\,mm flux density using available
data from the literature.

The source IRAS 19217+1651 was observed with the VLA by
\citet{Garay07}, who measured a 7\,mm flux density of $50.4 \pm0.16$
mJy towards the brighter of the two CARMA detections at 3\,mm. The 3mm
flux density of the brighter component is $66 \pm 1$ mJy; an additional
calibration uncertainty of $\sim$10\% should be included in both the
7\,mm and 3\,mm flux density measurements. The 7\,mm detection is
consistent with optically thin free-free emission \citep{Garay07}, and
implies that the free-free contribution to the 3\,mm flux density is
$\sim$46 mJy.  The HMPO IRAS 20126+4104 was observed with the VLA
between 20\,cm and 7\,mm by \citet{Hofner07}, who conclude that
free-free emission should contribute $\sim$4 mJy at 3\,mm, which is a
small fraction of the $33 \pm 2$ mJy detected with CARMA.  Our CARMA
observations detect three regions of 3\,mm emission towards the HMPO
IRAS 23033+5951, of which the northernmost is associated with 3.6\,cm
emission of 1.7 mJy \citep{Sridharan02}. For a ``typical'' UCHII radio
continuum spectrum, the free-free emission would contribute less than
4.6 mJy of the $8 \pm 1$ mJy detected with CARMA.  The HMPO IRAS
23139+5939 has a flux density of 1.4 mJy at 3.6\,cm
\citep{Sridharan02}, so we assume that the free-free emission
contributes no more than 3.8 mJy of the $12 \pm 2$ mJy detected with
CARMA at 3\,mm.  For regions without 3.6\,cm emission, we assume that
dust accounts for all of the observed 3\,mm flux density.

\subsection{Masses}

We derive the total (gas and dust) mass for each of the CARMA
continuum detections from the observed 3\,mm flux density, assuming
optically thin emission from isothermal dust, using the equation
\begin{equation} \label{MASSEQ}
M = \frac{S_\nu f_d D^2}{\kappa_\nu B_\nu(T)},
\end{equation}
where $S_\nu$ is the integrated flux density, $f_d$ is the fraction of
3\,mm emission from dust, $D$ is the distance, $\kappa_\nu$ is the
dust emissivity and $B_\nu(T)$ is the Planck function.  The value of
the dust emissivity is extrapolated from a value of $\kappa_{\rm o}=
0.1$\,cm$^2$~g$^{-1}$ at 250\,\micron\ assuming a power law spectral
index of $\beta = 2$ \citep[i.e. $\kappa_\nu = \kappa_{\rm o}
(\nu/\nu_{\rm o})^\beta$]{Hildebrand83}.  For 95\,GHz continuum
observations, the mass can be derived as
\begin{equation} \label{MASSAPPROX}
\frac{M}{\rm M_\odot} = 0.55 \left(\frac{S_\nu f_d}{\rm mJy}\right)
\left(\frac{D}{\rm kpc}\right)^2 \left(\frac{T}{\rm 50~K}\right)^{-1},
\end{equation}
where we use the Rayleigh-Jeans approximation to the Planck function
and assume a temperature of 50\,K.  Since the 3\,mm continuum emitting
regions are not resolved by our observations, a lower limit on the
volume density is derived assuming spherical geometry with a diameter
equal to the resolution and a mean molecular weight per free particle
of 2.37.  Estimated masses and densities for the individual 3\,mm
continuum sources are summarized in Table~\ref{OBSPROP}. The dominant
source of uncertainty in the derived masses is the dust emissivity,
which is uncertain by a factor of a few \citep{Hildebrand83}.

The masses of the sources detected in the CARMA 3\,mm continuum maps
range between 21 and 1200\solmass\ with volume densities in excess of
10$^8$\,cm$^{-3}$. These masses and densities can be compared to the
large scale core traced by single dish 1.2\,mm continuum maps
\citep{Beuther02a}. Assuming that the large scale core and the compact
clumps have the same temperature and dust emissivity,
we find that the compact 3\,mm
sources contain $\sim$20\% of the total core mass, and have volume
densities that are larger by a factor of $\sim$1000 than the average
core density.

\subsection{Outflows}

All 14 HMPO candidates in our sample were observed by
\citet{Sridharan02} with the CSO or IRAM 30~m telescopes to measure
the CO (2--1) line profile.  All were seen to have broad-line wings
consistent with bipolar outflows, suggesting that outflow activity is
common in these high-mass star forming regions.  The
\citet{Sridharan02} survey also searched for SiO (2--1) emission,
another common outflow tracer, from 12 of the 14 HMPO candidates in
our sample (all except IRAS 18517+0437 and IRAS 20332+4124), resulting
in seven detections.  Of the seven HMPO candidates with SiO (2--1)
emission, six are associated with 3~mm continuum emission detected by
CARMA, and none of the HMPO candidates without SiO (2--1) emission
were detected with CARMA.  Furthermore, of the five HMPO candidates
that have nearby 95~GHz \meth\ maser emission, four are also
associated with SiO (2--1) emission and the fifth was not included in
the survey of \citet{Sridharan02}.  However, the resolution of the
single-dish observations (10-30\arcsec) is not sufficient to determine
whether or not the HMPOs detected with CARMA are driving any these
outflows.

Interferometric observations have reported detections of outflows from
5 of the 14 HMPOs in our CARMA survey:

\noindent 1) \citet{Zhang07} observed SiO (2--1) at 5\arcsec\
resolution towards the HMPO IRAS 18566+0408, and found that the
emission traces a collimated outflow, originating from the position of
the CARMA 3~mm continuum detection.  

\noindent 2) SiO (2--1) and CO (2--1) emission from the HMPO IRAS
19217+1651 were observed by \citet{Beuther04} at 2-6\arcsec\
resolution.  There is a bipolar outflow seen in both transitions,
originating from the position of the CARMA 3~mm continuum source.

\noindent 3) IRAS 20126+4104 has been observed at high resolution by
\citet{Shepherd00}, who report the presence of a N-S oriented
molecular outflow and a NW-SE oriented jet centered on the position of
the CARMA continuum source.  Shocks identified by H$_2$ and [\sii]
emission trace the jet, and are coincident with the positions of the
95~GHz \meth\ masers detected by CARMA, in agreement with the
hypothesis that these maser lines are collisionally pumped at the
interface between jets/outflows and the ambient medium around the
protostar.

\noindent 4) CARMA found three 3~mm continuum peaks towards IRAS
23033+5951, one associated with 3.6~cm emission, one with a \water\
maser, and one with 95~GHz \meth\ maser emission.  This region has
been studied by \citet{Reid08}, who study the outflows in this region
at $\sim$6\arcsec\ resolution.  There are at least two outflows in
this region, one identified by HCO$^+$ (1--0) emission, and one
identified by \meth\ (2--1) and SiO (2--1) emission.  The CARMA 3~mm
continuum source coincident with the \water\ maser is most likely to
be driving the SiO outflow, and it is unclear what drives the HCO$^+$
outflow or is exciting the 95~GHz \meth\ maser.

\noindent 5) The 3~mm continuum source detected with CARMA in IRAS
23151+5912 is coincident with ``Peak 1'' in the 875 \micron\ continuum
map of \citet{Beuther07} and with the 3.4~mm and 1.3~mm continuum
source detected by \citet{Qui07}.  There are at least two outflows,
identified by SiO (2--1) and (8--7) emission, in this region, one
oriented roughly W-E and the other N-S \citep{Beuther07, Qui07}.  The
outflows are likely driven by two protostars, and it is not clear if
the 3~mm continuum emission detected with CARMA corresponds to one or
both protostars, or if the 95~GHz \meth\ maser is related to the
interaction between the W-E outflow and the ambient medium.

\section{Discussion} \label{DISCUSSION}

One proposed evolutionary sequence for HMPOs is that they originate
from low or intermediate mass protostars that are accreting material
from the surrounding core, presumably via a circumstellar disk
\citep{Beuther07}. As the protostar increases in mass, an \hii\ region
ultimately forms that eventually expands and halts accretion. While it
is unclear if \water\ or 6.7\,GHz \meth\ masers originate from the
disk or outflow, the observational evidence suggests these masers
trace an early stage of evolution before the \hii\ region has expanded
significantly \citep{Lo75, Genzel77, Walsh98, Codella04}. In this
scenario, a young HMPO with an accretion disk would be identified by
its water and/or methanol masers and mm-wavelength thermal emission
from the dust around the protostar.  When the HMPO is ``intermediate
aged'', an UCHII region detectable with cm-wavelength continuum
emission would grow but will not immediately disrupt the accretion, so
masers and dust thermal emission would also be present.  As the UCHII
region expands, it will destroy the disk and cut off the masers.  We
now evaluate whether or not the CARMA observations are consistent with
this evolutionary sequence.

First, we discuss if the compact sources detected in the CARMA 3\,mm
continuum maps signify the presence of a high mass protostar. The
masses of the compact sources range between 21 and 1200\solmass\
contained within a region \aboutless 10,000\,AU in diameter. Thus
these compact 3\,mm continuum sources are substantially denser and
more massive than the dense cores found in nearby clouds known to be
forming low mass stars. Even if we assume only 10\% of the core mass
forms into a star, a mid B-type star can potentially form in these
cores. Moreover, four of the 3\,mm continuum sources are spatially
coincident with compact \hii\ regions, confirming directly the
presence of a high mass star. Therefore, we assume that all of the
compact 3\,mm continuum sources represent dense dust and gas
surrounding a high mass star or protostar.  We cannot determine if the
material is distributed in a disk or compact envelope
\citep[e.g.][]{Beltran04,Cesaroni05,Reid08}, however, as the dust
continuum emission is unresolved by these observations.

Since 3\,mm continuum emission is detected toward 7 of the 14 HMPOs in
our survey we investigate whether the 50\% detection rate is set by
physical differences between the high mass cores or by selection bias.
A bias may be introduced since the HMPOs span a factor of $\sim$6 in
distance and a factor of $\sim$10 in luminosity, but the sensitivity
in each of the CARMA maps is approximately the same.  In the top panel
of Figure \ref{DISTNOISEPEAK}, we plot the noise in the CARMA 3\,mm
continuum maps as a function of the distance to the HMPO for sources
with (black circles) and without (gray circles) CARMA continuum
detections. No clear differences in the noise characteristics or
source distances are found between the two samples.  The middle panel
shows the core masses derived from single dish 1.2\,mm continuum maps
versus the bolometric luminosity of the core. Again, sources with and
without 3\,mm CARMA continuum detections share similar distributions,
indicating that the CARMA observations did not preferentially detect
cores with the highest mass or luminosity. Finally, the bottom panel
in Figure \ref{DISTNOISEPEAK} shows how the peak flux density in the
MAMBO single dish 1.2\,mm continuum maps varies with the integrated
1.2\,mm flux density. We detect (at the $\ge 3\sigma$ level) 3\,mm
continuum emission from all seven of the HMPOs with 1.2\,mm peak
fluxes above 300 mJy/(11\arcsec)$^2$, while none of the seven HMPOs
below that threshold are detected.

Is there an observational bias for detecting only those sources with
the highest peak surface brightnesses at 1.2\,mm, or is there an
intrinsic difference between those cores above and below the 300
mJy/beam threshold?  The peak 1.2\,mm flux densities in our sample are
all $\geq$ 100 mJy/beam.  If the 1.2\,mm flux density in the
11\arcsec\ single dish beam originates from an unresolved disk or
envelope, we would expect to observe flux densities of $\geq$4
mJy/beam at 3\,mm assuming an emissivity spectral index of $\beta =
1$, which is appropriate for disks \citep{Beckwith91}.  The noise in
the 3\,mm continuum maps is approximately 1 mJy/beam, and all 14 HMPO
candidates would be detectable at $\geq 3 \sigma$ level if the central
emission were compact.

Given that all 14 HMPO candidates could have been detected with CARMA,
we consider whether differences in the CARMA 3\,mm continuum detection
rate versus peak 1.2\,mm flux density could be explained by the
presence or absence of an unresolved disk or envelope. The median
integrated 3\,mm flux density detected with CARMA is 10\,mJy, which,
assuming an emissivity spectral index of $\beta = 1$, would contribute
$\sim$180\,mJy/beam of emission at 1.2\,mm.  This is approximately
equal to the difference between the peak 1.2\,mm flux densities of the
faintest HMPOs detected by CARMA ($\sim$ 400~mJy/beam @ 1.2\,mm) and
the brightest HMPOs not detected with CARMA ($\sim$ 250 mJy/beam).
Thus the lack of CARMA detections around sources with low peak 1.2\,mm
flux densities, but otherwise large integrated 1.2\,mm flux densities
and far-infrared luminosities, may indicate that a compact disk or
envelope has dissipated around these sources.

Since the detection rate of 3\,mm sources is not strongly biased by
the range of distances, masses and luminosities in our sample, we
investigate if the evolutionary stage correlates with the detection
rate of compact 3\,mm sources.  The signposts of the age sequence that
we are testing, in order of increasing age, are: (1) presence of
6.7\,GHz \meth\ and/or 22\,GHz \water\ maser emission and no cm
continuum emission, (2) both maser and cm continuum emission, and (3)
cm continuum emission but an absence of masers \citep{Lo75, Genzel77,
Codella94, Codella95}.  For each of the 14 massive cores in our
survey, we list the presence of 3\,mm continuum, cm continuum, and
maser emission in Table \ref{DETECTIONS}.

We detect 3\,mm continuum emission from 3/5 of the ``youngest'' HMPOs,
4/5 of the ``intermediate'' aged HMPOs and 0/4 of the ``oldest''
HMPOs.  Thus, the 3\,mm continuum emission is preferentially detected
toward the sources with maser emission, consistent with the proposed
evolutionary sequence for high mass stars. Further, we would expect
that the \water\ masers and the 6.7\,GHz \meth\ masers to be nearly
coincident with the compact 3\,mm sources, since VLBI observations
suggest that these masers originate from shocks associated with winds
or jets near the base of the molecular outflow \citep{Goddi05,
Moscadelli05}. All thirteen \water\ masers with interferometric
positions, and five of the eight 6.7\,GHz \meth\ masers, are in fact
located within 1.5\arcsec\ of a compact 3\,mm continuum source. It is
possible that the \meth\ maser sources without associated 3\,mm
continuum emission may have too low of disk or envelope mass to be
detected in our observations.

In summary, all of the \water\ and most of the 6.7\,GHz \meth\ maser
emission appear to be related to massive disks or envelopes, but these
disks/envelopes are not present toward sources that have radio
continuum emission but no maser emission.  The probability of
association between HMPOs with 3~mm continuum emission and HMPOs with
22~GHz \water\ masers and 6.7~GHz \meth\ masers is 98\% and 71\%,
respectively.  These 3~mm continuum sources are also likely to be
driving outflows, and there is a 98\% likelihood that 3~mm continuum
emission is associated with two possible outflow tracers, SiO (2--1)
emission and 95~GHz \meth\ masers.  There is a 99.8\% likelihood of
association between these two spectral lines.  These observations are
consistent with the hypothesis that as an \hii\ region around a high
mass protostar expands, it destroys the surrounding disk and envelope,
and consequently shuts off the source of the outflows and \water\ and
\meth\ masers.

\section{Summary} \label{SUMMARY}

In this paper we tested the hypothesis that \water\ and \meth\ masers
are signposts of an early phase in the evolution of a high mass
protostellar object before an expanding UCHII region has destroyed the
accretion disk.  We used CARMA to observe the 3\,mm continuum emission
at 1\arcsec\ resolution around a sample of HMPOs with and without
masers and cm continuum emission.  We also observed the 95\,GHz \meth\
line to determine its association with the thermal emission from dust
surrounding the protostars.

\begin{figure}
\epsscale{0.3}
\plotone{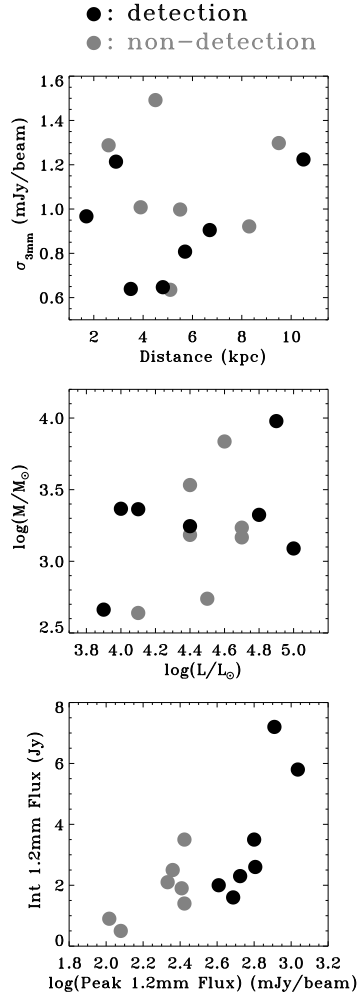}
\caption{Properties of the HMPO candidates with (black circles) and without
(gray circles) CARMA 3\,mm continuum detections.
{\it Top:} RMS noise in the CARMA 3\,mm image versus the distance to the 
HMPO. 
{\it Middle:} Mass of the dense core, as measured
by 1.2\,mm continuum images \citep{Beuther02a}, versus the bolometric
luminosity. 
{\it Bottom:} Integrated flux density in the 1.2\,mm single dish maps versus 
the peak 1.2\,mm flux density.
These figures show that HMPOs with and without CARMA detections have similar
noise characteristics, distances, core masses, and luminosities. However,
sources without CARMA continuum detections tend to have lower peak flux
densities. The data displayed in these plots are presented in
Tables~\ref{ARCHIVALTAB} and \ref{OBSPARAM}.
\label{DISTNOISEPEAK}}
\end{figure}

We detect (at $\ge 3\sigma$ significance) 3\,mm continuum emission
from all seven HMPOs with MAMBO 1.2\,mm continuum peak fluxes (taken
from Beuther et al.~2002a) above 300 mJy/(11\arcsec)$^2$, while none
of the seven HMPOs with peak fluxes below this threshold were detected
in our survey.  We argue that this is not merely an observational
selection effect, because all 14 HMPO candidates could have been
detected in the CARMA maps if their 1.2\,mm peak fluxes originated from
disks or compact envelopes, and were therefore concentrated into the
$\sim$1\arcsec\ CARMA beam.  The detection rate of HMPOs at 3\,mm is
not correlated with distance to the HMPO, the 1.2\,mm integrated flux 
detected with MAMBO, the total luminosity of
the HMPO or its mass (see Figure~\ref{DISTNOISEPEAK}).

All seven HMPOs detected at 3\,mm are associated with \water\ masers,
and only two of the seven HMPOs not detected at 3\,mm are associated
with \water\ masers.  The 22\,GHz \water\ masers and 95\,GHz \meth\
lines are significantly (98\% probability) associated with the
detection rate of 3\,mm continuum emission.  There is no significant
association between the detection rate of 3\,mm continuum and 6.7\,GHz
\meth\ or 1.7\,GHz OH masers or cm continuum emission.  The presence
of SiO (2--1) emission in the study of \citet{Sridharan02} was also a
useful predictor of CARMA 3~mm continuum detections (98\% probability)
and 95~GHz \meth\ masers (99.8\% probability).

The general picture of the evolution of high-mass protostars is
consistent with our observations that the 3\,mm continuum emission
comes from a disk and/or compact envelope around an HMPO.  Our
observations suggest that \water\ masers are associated with these
disks/envelopes, and that they are present both prior to and during
the formation of an UCHII region.  When the UCHII region disrupts the
disk and compact envelope around the HMPO, the \water\ and 95\,GHz
\meth\ masers are likely to be cut off and the 3\,mm continuum
emission falls below our threshold of detectability.

\acknowledgments

We thank the referee for improving the clarity and increasing the
scope of this paper.  We thank the CARMA staff, students and postdocs
for their help in making these observations.  We acknowledge support
from the Owens Valley Radio Observatory, which is supported by the
National Science Foundation through grant AST 05-40399.  Support for
CARMA construction was derived from the Gordon and Betty Moore
Foundation, the Kenneth T. and Eileen L. Norris Foundation, the
Associates of the California Institute of Technology, the states of
California, Illinois and Maryland, and the National Science
Foundation.  Ongoing CARMA development and operations are supported by
the National Science Foundation under a cooperative agreement, and by
the CARMA partner universities.  This research has made use of the
SIMBAD database, operated at CDS, Strasbourg, France.

{}

\clearpage
\begin{deluxetable}{rrrrrrrr} 
\tablewidth{0pt}
\tabletypesize{\scriptsize}
\tablecaption{Source Properties \label{ARCHIVALTAB}}
\tablehead{
 \colhead{{\it IRAS} Name}	                        & 
 \colhead{Group\tablenotemark{1}}                       &
 \colhead{distance\tablenotemark{2}}                    &
 \colhead{V$_{\rm LSR}$\tablenotemark{2}}               &
 \colhead{log(L$_{\rm total}$/\solum)\tablenotemark{2}} &
 \colhead{Peak $S_\nu$(1.2\,mm)\tablenotemark{3}}         &
 \colhead{Integrated $S_\nu$(1.2\,mm)\tablenotemark{3}}   &
 \colhead{Core Mass\tablenotemark{4}}                   \\
 \colhead{}                                             &
 \colhead{}                                             &
 \colhead{(kpc)}                                          &
 \colhead{(\kms)}                                         &
 \colhead{}                                             &
 \colhead{(mJy/beam)}                                     &
 \colhead{(Jy)}                                           &
 \colhead{(\solmass)}                                    
}
\startdata
18345$-$0641 & 1 &  9.5 &  95.9 & 4.6 & 265 & 1.4 & 6860\\
18440$-$0148 & 1 &  8.3 &  97.6 & 4.7 & 120 & 0.5 & 1717\\
18517+0437 & 1 &  2.9 &  43.9 & 4.1 & 812 & 7.2 & 2310\\
18566+0408 & 1 &  6.7 &  85.2 & 4.8 & 486 & 1.6 & 2110\\
23151+5912 & 1 &  5.7 & $-$54.4 & 5.0 & 406 & 2.0 & 1229\\
19217+1651 & 2 & 10.5 &   3.5 & 4.9 & 640 & 2.6 & 9518\\
20126+4104 & 2 &  1.7 &  $-$3.8 & 3.9 &  1087 & 5.8 & 460\\
20332+4124 & 2 &  3.9 &  $-$2.0 & 4.4 & 265 & 3.5 & 1529\\
23033+5951 & 2 &  3.5 & $-$53.1 & 4.0 & 631 & 3.5 & 2327\\
23139+5939 & 2 &  4.8 & $-$44.7 & 4.4 & 530 & 2.3 & 1759\\
19220+1432 & 3 &  5.5 &  68.8 & 4.4 & 256 & 1.9 & 3406\\
20205+3948 & 3 &  4.5 &  $-$1.7 & 4.5 &  104 & 0.9 & 548\\
22134+5834 & 3 &  2.6 & $-$18.3 & 4.1 &  229 & 2.5 & 436\\
22570+5912 & 3 &  5.1 & $-$46.7 & 4.7 &  215 & 2.1 & 1469\\
\enddata
\tablenotetext{1}{Group 1: With 22\,GHz \water\ and/or 6.7 \meth\ masers and no cm continuum, Group 2: With 22\,GHz \water\ and/or 6.7 \meth\ masers and cm continuum, Group 3: Without 22\,GHz \water\ or 6.7 \meth\ masers and with cm continuum}
\tablenotetext{2}{From \citet{Sridharan02}}
\tablenotetext{3}{1.2\,mm continuum flux densities from \citet{Beuther02a}, measured in a 11\arcsec\ beam}
\tablenotetext{4}{Mass of gas and dust inferred from 1.2\,mm continuum flux densities, taken from \citet{Beuther02a}}
\end{deluxetable}

\clearpage
\begin{deluxetable}{ccccccc} 
\tablewidth{0pt}
\tabletypesize{\scriptsize}
\tablecaption{CARMA Observing Parameters \label{OBSPARAM}}
\tablehead{
 \colhead{{\it IRAS} Name}	                & 
 \colhead{RA\tablenotemark{1}}	                &
 \colhead{Dec\tablenotemark{1}}                 & 
 \colhead{Phase Calibrators}                    &
 \colhead{Resolution\tablenotemark{2}}          &
 \colhead{Continuum Noise}                  	&
 \colhead{Noise/channel\tablenotemark{3}}       \\
 \colhead{}                                     &
 \colhead{J2000}                                &
 \colhead{J2000}                                &
 \colhead{}                                     &
 \colhead{($\arcsec\times\arcsec$)}             &
 \colhead{(mJy/beam)}                             &
 \colhead{(mJy/beam)}}
\startdata
18345$-$0641 & 18:37:17.02 & $-$06:38:30.70 & 1751+096,1743$-$038 & 1.6$\times$0.7 & 1.3 & 62 \\
18440$-$0148 & 18:46:36.56 & $-$01:45:21.20 & 1751+096,1743$-$038 & 1.6$\times$0.7 & 0.9 & 34 \\
18517+0437 & 18:54:14.32 & +04:41:39.69 & 1751+096,1743$-$038 & 1.4$\times$0.7 & 1.2 & 49 \\
18566+0408 & 18:59:10.02 & +04:12:14.69 & 1751+096,1743$-$038 & 1.6$\times$0.7 & 0.9 & 26 \\
19217+1651 & 19:23:58.77 & +16:57:44.80 & 1751+096,1925+211 & 1.5$\times$0.7 & 1.2 & 31 \\
19220+1432 & 19:24:20.05 & +14:38:03.60 & 1751+096,1925+211 & 1.4$\times$0.7 & 1.0 & 32 \\
20126+4104 & 20:14:25.86 & +41:13:33.99 & BLLAC,2007+404    & 1.4$\times$0.8 & 1.0 & 37 \\
20205+3948 & 20:22:20.87 & +39:58:15.00 & BLLAC,2007+404    & 1.6$\times$0.8 & 1.5 & 78 \\
20332+4124 & 20:34:59.72 & +41:34:49.40 & BLLAC,2007+404    & 1.2$\times$0.6 & 1.0 & 39 \\
22134+5834 & 22:15:09.51 & +58:49:05.99 & BLLAC,2007+404    & 1.5$\times$0.8 & 1.3 & 61 \\
22570+5912 & 22:59:05.37 & +59:28:19.18 & BLLAC,0102+584    & 1.0$\times$0.8 & 0.6 & 30 \\
23033+5951 & 23:05:25.31 & +60:08:06.28 & BLLAC,0102+584    & 1.0$\times$0.8 & 0.6 & 22 \\
23139+5939 & 23:16:10.45 & +59:55:28.48 & BLLAC,0102+584    & 1.0$\times$0.8 & 0.6 & 26 \\
23151+5912 & 23:17:21.02 & +59:28:48.48 & BLLAC,0102+584    & 1.2$\times$0.7 & 0.8 & 34 
\enddata
\tablenotetext{1}{Phase center positions taken from 1.2\,mm coordinates
listed in \citet{Beuther02a}}
\tablenotetext{2}{FWHM synthesized beam size with natural weighting}
\tablenotetext{3}{Channel width of 0.488 MHz}
\end{deluxetable}

\begin{deluxetable}{cccccrcrc} 
\tablewidth{0pt}
\tabletypesize{\scriptsize}
\tablecaption{Properties of Detected 3\,mm Continuum Sources \label{OBSPROP}}
\tablehead{
 \colhead{{\it IRAS} Name}	  & 
 \colhead{RA}                     &
 \colhead{Dec}                    & 
 \colhead{($\Delta \alpha, \Delta \delta$)\tablenotemark{1}} &
 \colhead{$S_{\rm peak}$}	  &
 \colhead{$S_{\rm int}$}          & 
 \colhead{$f_d$}\tablenotemark{2} &
 \colhead{Mass}                   &
 \colhead{Density}                \\
 \colhead{}                       &
 \colhead{J2000}                  &
 \colhead{J2000}                  &
 \colhead{($\arcsec,\arcsec$)}    &
 \colhead{(mJy/beam)}             &
 \colhead{(mJy)}                  &
 \colhead{}                       &
 \colhead{(\solmass)}             &
 \colhead{(10$^8$ cm$^{-3}$})}
\startdata
18517+0437 & 18:54:14.24 & +04:41:40.7 & ($-$1.1, 1.0)   & 8.5$\pm$1.4 & 16$\pm$2 & 1.0 &   74 &  8.6 \\ 
18566+0408 & 18:59:09.99 & +04:12:15.3 & ($-$0.4, 0.6)   & 4.4$\pm$1.2 & 10$\pm$2 & 1.0 &  250 &  2.3 \\ 
19217+1651 & 19:23:58.81 & +16:57:41.1 & ( 0.6,$-$3.7)   & 55$\pm$3    & 66$\pm$1 & 0.3 & 1200 &  2.9 \\ 
19217+1651 & 19:23:58.70 & +16:57:42.0 & ($-$1.1,$-$2.8) & 4.9$\pm$1.3 &  9$\pm$2 & 1.0 &  540 &  1.3 \\ 
20126+4104 & 20:14:26.03 & +41:13:32.6 & ( 1.8,$-$1.4)   & 18$\pm$1    & 33$\pm$2 & 0.9 &   47 &  27  \\ 
23033+5951 & 23:05:24.65 & +60:08:09.3 & ($-$4.9, 3.0)   & 4.4$\pm$0.9 &  8$\pm$1 & 1.0 &   54 &  3.6 \\ 
23033+5951 & 23:05:24.97 & +60:08:14.2 & ($-$2.5, 7.9)   & 2.3$\pm$0.6 &  6$\pm$1 & 1.0 &   40 &  2.7 \\ 
23033+5951 & 23:05:25.09 & +60:08:16.3 & ($-$1.7,10.0)   & 3.4$\pm$0.9 &  8$\pm$1 & 0.4 &   21 &  1.4 \\ 
23139+5939 & 23:16:10.45 & +59:55:28.6 & ( 0.0, 0.1)     & 2.8$\pm$0.5 & 12$\pm$2 & 0.7 &  110 &  2.7 \\ 
23151+5912 & 23:17:20.94 & +59:28:47.6 & ($-$0.8,$-$0.9) & 3.5$\pm$1.2 &  8$\pm$2 & 1.0 &  140 &  2.2 \\ 
\enddata
\tablenotetext{1}{Difference between the measured 3\,mm continuum source and the 1.2\,mm core (see Table~\ref{ARCHIVALTAB})}
\tablenotetext{2}{Fraction of 3\,mm emission from dust (see Section \ref{RESULTS})}
\end{deluxetable}

\begin{deluxetable}{lcccccc} 
\tablewidth{0pt}
\tabletypesize{\scriptsize}
\tablecaption{Properties of Detected 95\,GHz \meth\ Spectral Lines\label{METHPROP}}
\tablehead{
 \colhead{{\it IRAS} Name}             & 
 \colhead{RA}                          &
 \colhead{Dec}                         & 
 \colhead{separation\tablenotemark{1}} &
 \colhead{$S_{\rm peak}$}	       &
 \colhead{V$_{\rm LSR}$}               & 
 \colhead{FWHM}                        \\
 \colhead{}                            &
 \colhead{J2000}                       &
 \colhead{J2000}                       &
 \colhead{(10$^4$\,AU)}                   &
 \colhead{(Jy)}                          &
 \colhead{(\kms)}                        &
 \colhead{(\kms)}}
\startdata
18517+0437 (a) & 18:54:14.74 & +04:41:42.6 & 2.22$\pm$0.05 & 2.98 &    43.4 & 1.2 \\
18517+0437 (b) & 18:54:14.46 & +04:41:44.6 & 1.48$\pm$0.05 & 1.60 &    43.9 & 1.4 \\
20126+4104 (a) & 20:14:26.72 & +41:13:29.8 & 1.40$\pm$0.01 & 2.72 &  $-$4.2 & 1.5 \\
20126+4104 (b) & 20:14:25.24 & +41:13:34.9 & 1.57$\pm$0.01 & 1.46 &  $-$2.1 & 2.2 \\
20126+4104 (c) & 20:14:25.16 & +41:13:36.4 & 1.79$\pm$0.01 & 2.92 &  $-$2.4 & 2.0 \\
20126+4104 (d) & 20:14:25.44 & +41:13:37.3 & 1.39$\pm$0.01 & 6.14 &  $-$2.9 & 1.4 \\
20126+4104 (e) & 20:14:25.41 & +41:13:37.9 & 1.49$\pm$0.01 & 6.35 &  $-$2.9 & 1.3 \\
23033+5951     & 23:05:24.59 & +60:08:09.4 & 0.16$\pm$0.04 & 4.55 & $-$54.3 & 1.4 \\
23139+5939 (a) & 23:16:10.86 & +59:55:20.7 & 4.07$\pm$0.13 & 0.44 & $-$44.6 & 1.7 \\
23139+5939 (b) & 23:16:10.40 & +59:55:28.0 & 0.33$\pm$0.13 & 1.23 & $-$45.9 & 1.8 \\
23151+5912     & 23:17:21.90 & +59:28:45.7 & 4.30$\pm$0.22 & 1.87 & $-$52.8 & 1.4 
\enddata
\tablenotetext{1}{Projected separation between 95\,GHz \meth\ maser
position and nearest 3\,mm continuum peak. Uncertainties estimated
from signal to noise of the maser and continuum detections.}
\end{deluxetable}

\begin{deluxetable}{cccccccc} 
\tablewidth{0pt}
\tabletypesize{\scriptsize}
\tablecaption{Association of Masers and Continuum Detections\label{DETECTIONS}}
\tablehead{
 \colhead{{\it IRAS} Name}	              & 
 \colhead{Group\tablenotemark{1}}             &
 \colhead{3\,mm Continuum\tablenotemark{2}}   &
 \colhead{3.6\,cm Continuum\tablenotemark{3}} &
 \colhead{\water\ Maser\tablenotemark{3}}     &
 \colhead{\meth\ Maser\tablenotemark{3}}      &
 \colhead{\meth\ Maser\tablenotemark{2}}      &
 \colhead{OH Maser\tablenotemark{4}}          \\
 \colhead{}		                      & 
 \colhead{}		                      & 
 \colhead{}	                              &
 \colhead{}                                   & 
 \colhead{22\,GHz}                            & 
 \colhead{6.7\,GHz}	  	              & 
 \colhead{95\,GHz}                            &
 \colhead{1.7\,GHz}}
\startdata
18345$-$0641 & 1 & N & N & Y & Y & N & Y \\
18440$-$0148 & 1 & N & N & N & Y & N & Y \\
18517+0437   & 1 & Y & N & Y & Y & Y & N \\
18566+0408   & 1 & Y & N & Y & Y & N & Y \\
23151+5912   & 1 & Y & N & Y & N & Y & N \\
19217+1651   & 2 & Y & Y & Y & Y & N & Y \\
20126+4104   & 2 & Y & Y & Y & Y & Y & Y \\
20332+4124   & 2 & N & Y & Y & N & N & N \\
23033+5951   & 2 & Y & Y & Y & N & Y & N \\
23139+5939   & 2 & Y & Y & Y & Y & Y & Y \\
19220+1432   & 3 & N & Y & N & N & N & Y \\
20205+3948   & 3 & N & Y & N & N & N & N \\
22134+5834   & 3 & N & Y & N & N & N & N \\
22570+5912   & 3 & N & Y & N & N & N & N \\
\enddata
\tablenotetext{1}{Group 1: With 22\,GHz \water\ and/or 6.7 \meth\ masers and no cm continuum, 
                  Group 2: With 22\,GHz \water\ and/or 6.7 \meth\ masers and cm continuum, 
                  Group 3: Without 22\,GHz \water\ or 6.7 \meth\ masers and with cm continuum}
\tablenotetext{2}{This study}
\tablenotetext{3}{\citet{Sridharan02}}
\tablenotetext{4}{\citet{Edris07}}
\end{deluxetable}


\end{document}